\documentstyle[11pt,newpasp,twoside,epsf]{article}
\def\edcomment#1{\iffalse\marginpar{\raggedright\sl#1\/}\else\relax\fi}
\reversemarginpar

\begin{document}
\title{Light-weight Local Group Dwarf Spheroidal Galaxies}
 \author{J. R. Kuhn and D. Kocevski}
\affil{Institute for Astronomy, Univ. of Hawaii, 2680 Woodlawn Dr., Honolulu, HI, 96822, USA}
\author{J-J Fleck}
\affil{D\' epartement de Physique de l'\' Ecole Normale Sup\' erieure, 24 rue 
Lhomond, F-75231 Paris, Cedex 05, France}

\begin{abstract} 
A simple and natural explanation for the dynamics and
morphology of the Local Group Dwarf Spheroidal galaxies, Draco (Dra) and
Ursa Minor (UMi), is that they are weakly unbound stellar systems with
no significant dark matter component.  A gentle, but persistent, Milky
Way (MW) tide has left them in their current kinematic and
morphological state. This short paper reviews the parametric tidal
interaction model which accounts for their behavior and discusses new
statistical evidence from the observed stellar distribution in Dra
which implies that its total mass is not dominated by collisionless
dark matter (DM).  
\end{abstract}

\section{Introduction}

The Milky Way may be our best testing ground for models of
cosmologically important dark matter masses. The clumpiness of the MW
potential and the mass and number density of dwarf (M$<10^8M_\odot$)
galaxies is an important constraint on ``standard" Cold Dark Matter
(CDM) mass components (Mayer et al. 2002). The paucity of dwarf galaxies in the Local
Group is something of a crisis for CDM (Klypin et al. 1999) and adds to the urgency
of understanding what the masses of the MW dS really are.

Controversy over the dynamical masses of the MW dS centers around
whether or not they are in equilibrium. At one extreme the equilibrium
models account for  elliptical morphology and dynamics  with ad
hoc multiparametric dark matter mass distributions (cf. Kleyna et al. 2002). In contrast a
variety of non-equilibrium dS models account for their properties
through MW tidal interactions (Kuhn and Miller 1989; Kroupa 1997).  Several years ago Kuhn and
Miller (1989) argued that if only the dS luminous mass component is
present (as is the case in globular clusters) then their dynamical
timescales are comparable to their MW orbital times. This has important
consequences  since then even a relatively weak MW tidal force can
profoundly affect the dS through its orbital time dependence. Recently
we developed analytic and numeric models for a parametric resonance
that describes how a variable tidal interaction inflates the velocity
dispersion of a dS stellar system (Fleck and Kuhn 2003 - FK).

In a parallel empirical effort we have been looking for tidal tails as
evidence of dS non-equilibrium. The Sagitarrius dS tail is the most
obvious example of a dS-MW tidal interraction which, incidently, was
discovered (Ibata et al. 1994) after the early non-equilibrium tidal predictions. Other
examples of tails beyond a dS tidal radius and with a surface
density far above what is expected from a massive equilibrium system
have been discovered in Carina (Kuhn et al. 1996), Ursa Minor (Smith et al.
1997), and Sculptor (Walcher et al. 2003).

Here we also briefly describe a new test of the simplest equilibrium
dark matter hypothesis in dS using relatively deep color-magnitude
photometry from our CFHT wide-field survey of Ursa Minor and Draco. By
color-selecting and then counting stars into the dS main sequence the
statistical distribution of the stellar number density becomes a
sensitive probe of the overall dS gravitational potential.

\section{Parametric Tidal Oscillations}

The cumulative effect of a weak MW tide acting on a dS has been
described by FK in terms of a modified two-dimensional oscillator.
This equation is analogous to a one-dimensional Mathieu
equation (cf. Bender and Orszag 1978) 
which describes an inverted pendulum or the amplitude
of a ``pumped'' swing oscillator.

\begin{eqnarray}\label{eq:xy}
\ddot{x} + {\omega_0}^2 ( {1 - \varepsilon\, \cos{2\varphi}) x	}	
		& = &	\varepsilon\, {\omega_0}^2\, y\sin{2\varphi}		\\
\ddot{y} + {\omega_0}^2 ( {1 + \varepsilon\, \cos{2\varphi}) y	}	
		& = &	\varepsilon\, {\omega_0}^2\, x\sin{2\varphi} 
\end{eqnarray}

The dS-centered local cartesian coordinates describe the position
of a star and $\varphi$ is the orbital phase angle of the dS in its MW
orbit. Here $\varepsilon = {\omega_c^2/\omega_0^2}$ is the strength of the
tide expressed in terms of the ratio of the MW circular frequency and
the harmonic frequency of the self-gravitating dS.
Solutions to this equation are known as parametric oscillations
and have several interesting properties.

Unlike driven simple harmonic oscillations, where the growth rate of the
oscillation depends strongly on the intrinsic mode damping, parametric
modes can exhibit secular growth even when the intrinsic harmonic
oscillation frequency is significantly different from the driving
frequency.  If the orbital phase of the dS is $\varphi =\omega t$ then FK
showed that 2-d
parametric modes are unstable when $\omega_0(1-\varepsilon /4) < \omega <
\omega_0(1+\varepsilon /4)$.  Non-harmonic dS potentials exhibit qualitatively
similar behavior.  FK found that a large fraction of possible dS orbits
are subject to parametric excitation. They also showed that parametric
oscillations generally cause an inclined dS bar with an
inclination angle that depends on the dS-MW orbit and potential, and with
a growth rate which is a simple function of $\varepsilon$.

From numeric and analytic solutions to the tide equations FK showed
that centrally concentrated dS stellar distributions, but with inflated
velocity dispersions, could be generated after several MW orbits. In
general the tidally induced velocity dispersion of any dS system should
increase with $\varepsilon$. This dependence is consistent with the
observed MW dS. Figure 1 shows how the ``apparent M/L'' varies with the
empirical mode growth estimate ($A=\sqrt{\varepsilon}$) based on a
standard logarithmic MW potential and the luminous mass of each of the
dS. The large apparent $M/L$ ratio of each of the local dS is
consistent with this model of parametric tidal excitation, leaving no
kinematic evidence for a significant dark matter binding mass.

\begin{figure}
\plotfiddle{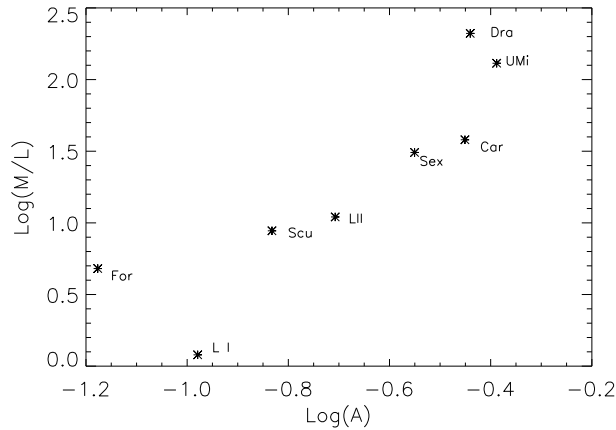}{2in}{0}{50}{50}{-170}{-190}
\caption{Variation of dS apparent M/L with parametric mode growth estimate.}
\end{figure}

\section{Stellar Number Density Statistics}

There exist equilibrium mass calculations for Dra which claim a M/L in
excess of 500 solar units (Kleyna et al. 2002). With this assumption the Dra
potential must be dominated by its dark matter. If we assume the DM is
collisionless and relaxed, and that it and the baryonic mass of the dS
formed at the same epoch, then the dS stars are, effectively, massless
test particles within a smooth DM-dominated potential.  Such stellar
``particles'' must be uncorrelated, with a null two-particle
correlation function (cf. Binney and Tremaine 1987). It follows that the stellar
particle distribution function, $f(\vec r,\vec v,t)$, can be computed
from the collionless Boltzmann equation and the net dS gravitational
potential (cf. Binney and Tremaine 1987).

For a smooth gravitational potential we also expect $f(\vec r)$ to be smooth
so that the projected star counts in a small spatial region (for
example as derived from one pixel of a two-dimensional star count map)
can yield a reasonable estimate of the projected number density, $n(\vec x)$.
Here $\vec x$ is a 2-d coordinate in the plane of the sky and a line-of-sight and velocity integration of $f(\vec r, \vec v, t)$ to obtain $n(\vec x)$
is implicit. On scales small compared to the variation of the potential
we also expect $n(\vec x)$ to be smooth. In fact, since stars are
uncorrelated, our estimate of $n(\vec x)$ from the star counts per pixel should be Poisson distributed
-- like the photon count in a 2-dimensional image. Any spatial
``clumpiness'' in the number of stars per pixel which is in excess of
Poisson distribution variations violates our assumption that Dra has a smooth DM
dominated potential.

Using deep V and I color-magnitude data obtained from the CFHT 12K CCD
mosaic camera over a $2\times 1\deg$  region centered on Dra
(Kocevski et al. 2003) we estimate the circularly symmetric projected
number density $n(R)$ when $R=\|\vec x\|$ (Figure 2). The effect of non-dS
background stars was minimized by counting stars in a color-magnitude
region that included the Dra giant branch, horizontal branch and main
sequence down to V=25. Radial bins were
scaled along the y axis to account for the ellipticity of Dra. The
radial binwidth also increased with $R$ in order to maintain a constant
count per bin, and constant statistical significance, since the mean density
$n(R)$ decreased with $R$. 

\begin{figure}
\plotfiddle{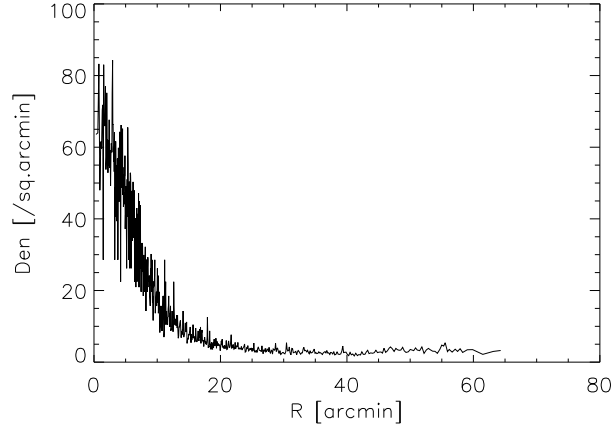}{2in}{0}{50}{50}{-170}{-190}
\caption{The number density of Dra stars in elliptical bins versus projected
distance from the Dra center.}
\end{figure}

Figure 2 shows $n(R)$ in approximately $10^3$ radial bins from a region
slightly larger than $1 \deg$ in radius centered around Dra. The
fluctuations in $n(R)$ in groups of 5 bins were used to compute the
normalized (reduced) chisquared. At radial bin $R_i$ we obtain (with
the assumption of Poisson statistics)
  
$$\chi^2_5(R_i)={1\over 5}\sum_{j=0}^4\sigma^2(R_{i+j})/n(R_{i+j})$$
Here $\sigma^2(R_i)$ is the sample variance computed at each
radius bin from it and 4 nearby bins.  We find that $\chi^2$ derived from
the outer bins (with $R > 20\arcmin$) is consistent with the expected reduced $\chi^2$
distribution. The inner Dra bins have a variance which is too large.
Figure 3 plots the distribution of $\chi^2$ measured this way for the
inner   Dra region. Using a Kolmogorov-Smirnoff test we find that
the inner
region is not consistent with the expected $\chi^2_5$ distribution at
better than a 99\% confidence level (while the
outer region is). Thus, the stellar density in the inner
region of Dra is ``clumpier'' and more variable with spatial position than
it should be if the Dra mass were dominated by a smooth DM component.

\begin{figure}
\plotfiddle{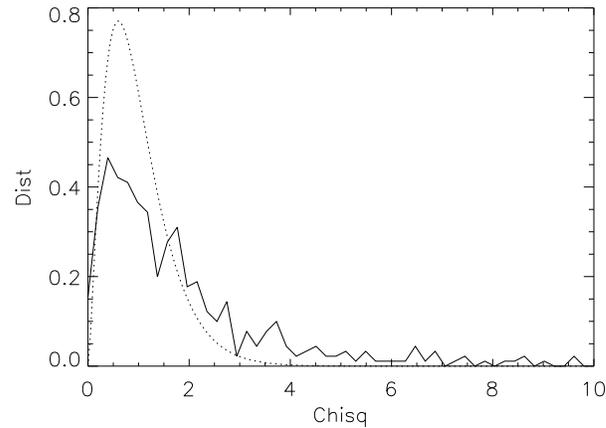}{2in}{0}{50}{50}{-170}{-190}
\caption{The actual (solid) and expected (dotted) $\chi^2_5$ for the
stellar number density variance in the inner $20 \arcmin$ of Dra is plotted here.
The preponderance of larger-than-expected empirical $\chi^2_5$ values implies
that the potential of Dra is not dominated by a smooth DM mass distribution.}
\end{figure}

\end{document}